\documentclass[%preprint, 
 amsmath,amssymb,
 aps, physrev,
]{revtex4-2}

\usepackage{graphicx}% Include figure files
\usepackage{dcolumn}% Align table columns on decimal point
\usepackage{bm}% bold math
\usepackage{xcolor}
\usepackage{xspace}
\usepackage{natbib,hyperref}
\begin{document}

\preprint{APS/123-QED}

%\title{\textbf{Thermo-Hydrodynamics dictate the interface crossing behavior of a prolate microswimmer} 
\title{Interface crossing behavior of prolate microswimmers: thermo and hydrodynamics}
%}% 

\author{Rishish Mishra}
\author{Harish Pothukuchi}%
\affiliation{%
 Department of Mechanical Engineering, Indian Institute of Technology Jammu
181221, India.
}%

\author{Harinadha Gidituri}
 \email{g.harinadha@hyderabad.bits-pilani.ac.in}
\affiliation{
 Department of Chemical Engineering, Birla Institute of Technology Pilani, Hyderabad
Campus 500078, India
}%

\author{Juho Lintuvuori}
 \email{juho.lintuvuori@u-bordeaux.fr}
\affiliation{%
 Univ. Bordeaux, CNRS, LOMA, UMR 5798, F-33400 Talence, France
  \\
}%

\date{\today}% It is always \today, today,

\begin{abstract}
Motivated by recent experiments of motile bacteria crossing liquid-liquid interfaces of isotropic-nematic coexistence (Cheon {\it et al.}, Soft Matter 20: 7313-7320, 2024), we study the dynamics of prolate microswimmers traversing clean liquid-liquid interfaces. Using large-scale lattice Boltzmann simulations, we observe that neutrally wetting swimmers can be either trapped or cross the interface, depending on their initial angle, swimming speed and the interfacial tension between the two fluids. The simulation results are rationalized by considering a competition between interfacial (thermodynamic) and active (hydrodynamic) forces. The swimmers get trapped at the interface due to a thermodynamic trapping force, akin to Pickering effect, when the forces from interfacial tension dominate over the swimming forces. The trapping behavior can be captured by calculating a critical capillary number by balancing the interfacial and active energies. This prediction agrees remarkably well with the numerical simulations as well as the bacterial experiments of Cheon {\it et al.}, (Soft Matter 20: 7313-7320, 2024). Finally, our results demonstrate that the torque resulting in a reorientation of the swimmers parallel to the interface have both hydro and thermodynamic components. % 

\end{abstract}
\maketitle

\section{\label{sec:level1}Introduction\protect }

The motility of microorganisms in an aqueous environments is important for their evolution. 
The locomotion helps them in seeking nutrient-rich hot spots, for reproduction, and to escape from a protist predators. In particular, their motion strongly depends on the feedback mechanism from the surrounding environment \cite{ELGERTI_ROP_2015}. Such feedback can result from any source, for example, another particle (passive or active) in the vicinity or a surface such as a wall or fluid-fluid interface. It is well known from the previous experimental \cite{DENG_JFM_2022,Cheon_SOFT_MATTTER_2024,Cheon2024_Arxiv_2024} and numerical studies \cite{GIDITURI_PRF_2022,FENG_2022_PRR,FENG_2023_POF} that hydrodynamics dictates movement of microorganisms such as bacteria when swimming close to a no-slip wall or to fluid-fluid interfaces. Specifically, during the interaction with the fluid-fluid interface, in addition to hydrodynamics, thermodynamics also modifies the dynamic behavior of micro-particles (passive and active) ~\cite{GIDITURI_POF_2021}. Sometimes these microswimmers gets adsorbed from bulk onto the interface and straddle upon it. {\color{black} For neutrally wetting colloidal particles,} thermodynamics drives such an adsorption process by reducing the available interfacial area which in turn reduces the interfacial energy (Pickering effect)~\cite{VACCARI_ACIS_2017,WILLIAMS_JCIS_1992,PIERANSKI_PRL_1980,DU_LANGMUIR_2010}.
        
The interaction between microswimmer and interface has recently attracted significant interest due to its applications, including pellicle formation \cite{DESAI_SOFTM_2020}, biomedical applications \cite{LI_ACS_2016}, and environmental remediation \cite{GAO_RSC_2013, SOLER_ACS_2013,JURODO_SMALL}. Motivated by this, several researchers have explored the interactions of microswimmers at and near the interface, encompassing experimental \cite{Deng_Langmuir_2020, DENG_JFM_2022,Cheon_SOFT_MATTTER_2024,Cheon2024_Arxiv_2024}, theoretical \cite{CHISHOLM_JFM_2021,MENZEL_JCP_2016,SHAIK_JFM_2017}, and numerical approaches\cite{ISHIMOTO_PRE_2013,MALGARETTI_SOFT_2016,Feng_PRR_2022,FENG_2023_POF,RISHISH_FDR_2024}. For instance, Shaik and Ardekani \cite{SHAIK_JFM_2017} studied the hydrodynamic behavior of a microswimmer near a weakly deformable interface. Using a time-reversible squirmer, they reported swimmer either moves towards or away from the interface, depending upon its initial velocity and orientation. Similarly, Desai \textit{et al.} \cite{DESAI_SOFTMATTER_2018} investigated the hydrodynamics of microswimmers modeled as a force dipole near a drop and found that the swimmers orbit around or scatter away subject to the drop radius (critical trapping radius) and basin of attraction. They concluded that influence of hydrodynamic attraction does not extend beyond three body length of the swimmer. Further, Feng \textit{et al.} numerically investigated the dynamics of a spherical microswimmer near a liquid-liquid interface with \cite{FENG_2023_POF} and without \cite{FENG_2022_PRR} the viscosity contrast. They observed different swimming states such as, bouncing, penetrating, hovering, and sliding over the interface. These states are presented as a function of incident angle and microswimmer type. Given that microswimmers can exhibit a variety of shapes, Nambiar and Wettlaufer \cite{NAMBIAR_PRF_2022} studied the hydrodynamics of a slender shaped microswimmer near a deformable fluid-fluid interface. Their results reveal the mutual interaction (attraction and repulsion) between the swimmer and the interface (deformation), governed by the swimmer's configuration, position, and type (pusher or puller). Later, they also analyzed the role of different type of stochasticities (random tumble and active Brownian) on the reorientation behavior of the microswimmer \cite{NAMBIAR_PRF_2024}. Their findings revealed that both rotational and positional shifts occurred, either toward or away from the interface. 

{\color{black} While these studies have clearly analyzed the hydrodynamics interaction between the microswimmer and the fluid-fluid interface, however the role of the interfacial trapping (Pickering) forces is often neglected.} 
 A recent experimental study, {\color{black} Cheon} \textit{et al.}~\cite{Cheon_SOFT_MATTTER_2024} examined the interaction between a rod-shaped bacteria \textit{Bacillius subtilis} with an liquid-liquid {\color{black} interface consisting of a coexisting isotropic-nematic phase.  The results demonstrated that bacteria can either be trapped or cross the interface depending on their} velocity and initial projection angle.  The trapping results were rationalized considering an interfacial force and torque that arise from an interfacial deformation created by the bacteria pushing on the interface between the isotropic and nematic states~\cite{Cheon_SOFT_MATTTER_2024}. 
  \textcolor{black}{Further, the competition between the interfacial and active forces have been observed to lead to a repartition of \textit{Bacillius subtilis} in binary fluids consisting of dextran (DEX) and polyethylene glycol (PEG) interfaces~\cite{Cheon2024_Arxiv_2024}.} 

 Motivated by these experiments, we study the trapping dynamics of a rod-like microswimmers at clean liquid-liquid interfaces. We demonstrate that the trapping dynamics of neutrally wetting swimmers is dominated by the Pickering effect. It can be captured by considering a capillary number based on the ratio between swimming force and {\color{black} and a force arising from the interfacial tension $\sigma_0$ between the two fluids: $\mathrm{Ca}=u_0\eta/\sigma_0$, where $u_0$ is the swimming speed of the bacteria and $\mathrm{\eta}$ is the fluid viscosity.} When the thermodynamic forces {\color{black} arising from the interfacial tension} dominate, the swimmer is trapped. Further, we show that the rotational dynamics of the swimmers, arises from combined thermo and hydrodynamic torques. % For typical parameters, the reorienting torque arising from the force-dipole flow-field is observed to dominate over the passive, interfacial tension driven torque.}

{\color{black} To do this, we use } a lattice Boltzmann Method \cite{Desplat_CPC_2001,Kendon_JFM_2001} to model the interaction of a prolate microswimmer approaching an ideal fluid-fluid interface at varying capillary numbers $\mathrm{Ca}$ and projection angles $\phi_0$. We establish a relation between them using an energy balance equation to predict the interface crossing behavior (trapped or crossing) of the microswimmers.  Additionally, we also discuss the hydrodynamic rotation of a microswimmer at the interface. {\color{black} The predictions and simulations of the trapping, agree remarkably well with the experimental observations of bacteria at isotropic-nematic coexistence phase~\cite{Cheon_SOFT_MATTTER_2024}, even though the proposed physical mechanism is different.}

\subsection{\label{sec:level2}Simulation Model and details} \label{Simulation details}

We have performed 3D simulations for the geometry of a prolate-shaped microswimmer described by the equation $\frac{x'^2}{a^2}+\frac{y'^2}{b^2}+\frac{z'^2}{c^2} = 1$ where, $a>b=c$, with $a$ and $b$ being the semi-major and semi-minor axes, respectively. The aspect ratio of the prolate is defined as $AR= \frac{a}{b}$. In the present work, we fix the value of $AR$ at 3 with $b = 4$. 

Hydrodynamics of the prolate microswimmer under Stokes flow is captured using the spheroidal squirmer model \cite{THEERS_SOFTMATTER_2016}. In this scheme, a microswimmer is considered as a solid prolate particle with a prescribed slip velocity on the boundary, defined as following
\begin{equation}
    v_s = -B_1 (\textbf{s}.\hat{e})\textbf{s} - B_2 \zeta(\textbf{s}.\hat{e})\textbf{s}
\end{equation}
where $\hat{e}$ is along the major axis. The surface tangent vector $\textbf{s}$ is given as,

\begin{equation*}
    \textbf{s} = - \dfrac{\sqrt{a^2-z'^2}}{\sqrt{a^2-\epsilon^2z'^2}}\hat{e} + \dfrac{\sqrt{1-\epsilon^2}}{\sqrt{a^2 - \epsilon^2z'^2}}z'\hat{e}_\perp
\end{equation*}
when the major axis of prolate particle $\hat{e}$ is oriented along $z'$ axis\cite{KELLER_JFM_1977}. The spheroidal coordinate system term $\zeta = [\sqrt{x'^2 + y'^2 +(z'+a\epsilon)^2}-\sqrt{x'^2+y'^2+(z'-a\epsilon)^2}]/2a\epsilon$.

The amplitudes of the slip velocity, $B_1$ and $B_2$ are the first two squirming modes, representing the source and force dipoles, respectively. The ratio of these two squirming modes, $\dfrac{B_2}{B_1}$ is called the squirmer coefficient, denoted by $\beta$. This coefficient can be varied to model different types of microswimmers such as pusher ($\mathrm{\beta < 0} $), puller ($\beta > 0 $) and neutral ($\beta = 0 $). However, in the current study, we kept the $\beta$ value at -1. Such a choice is motivated from the experiments ~\cite{Cheon_SOFT_MATTTER_2024} on Bacteria (\textit{Bacillus subtilis}) which is of a pusher type. 
 The swimming speed $u_0$ for a prolate microswimmer 
$$u_0=B_1\epsilon^{-1}(\epsilon^{-1}-(\epsilon^{-2}-1)\coth^{-1}(\epsilon^{-1})),$$
where $\epsilon = \sqrt{1-\frac{1}{AR^2}}$\cite{THAMPI_PRE_2024}. 

\begin{figure}
\centering
  \includegraphics[width=0.5\textwidth]{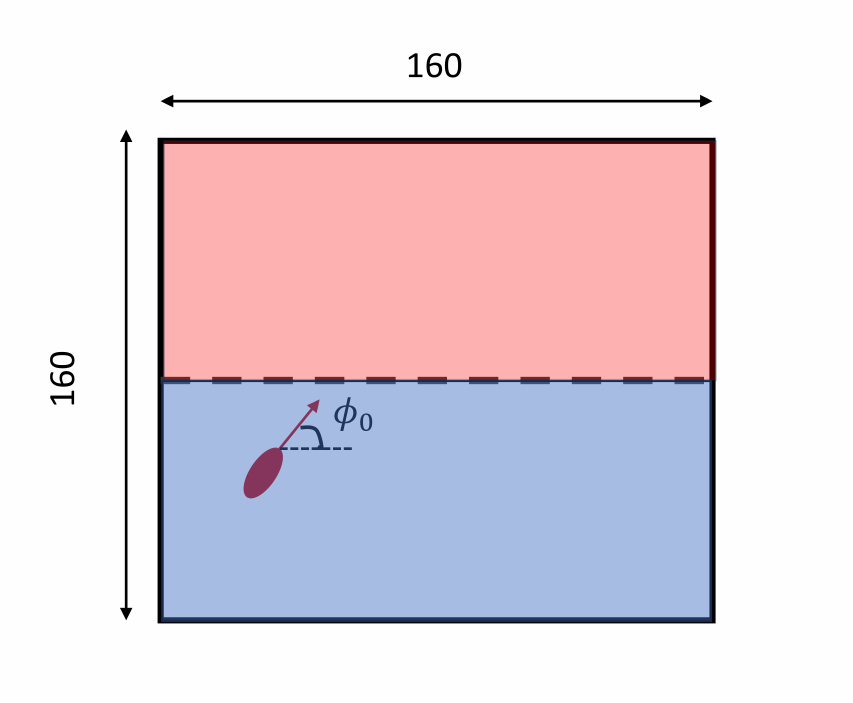}
  \caption{A prolate microswimmer approaching towards interface separating two fluids with self-propulsion velocity $u_0$ and initial projection angle $\phi_0$.}
  \label{schematic}
\end{figure}

Next, a cubical simulation box with dimensions $160 \times 160 \times 160$ is considered, where the interface is positioned at the center of the box. The box has periodic boundaries in two directions, while solid wall boundaries are applied along the plane parallel to the interface. The projection angle $\phi_0$ is defined as the angle between the interface and the microswimmer orientation ($\hat{e}$) along the major-axis  (Fig. \ref{schematic}). 

The existence of an interface is due to the presence of two immiscible fluids. To model the interface of a binary fluid mixture, Ginzburg-Landau type free energy functional is used \citep{Kendon_JFM_2001} and is given as below.

\begin{equation} \label{eq:free_energy}
  F =  \int dV \left( -\frac{A}{2}\psi^2+ \frac{B}{4}\psi^4+\frac{\kappa}{2}|\nabla{\psi}|^2\right)   
\end{equation}

\noindent Here $\mathrm{\psi}$ is the phase field variable. The temporal evolution of $\mathrm{c}$ is governed by the Cahn-Hilliard equation,

\begin{equation} \label{eq:chan-hillard}
\partial_{t} \psi + {\bf{U}}\cdot \nabla \psi = M \nabla^2 \mu
\end{equation}

\noindent where ${\bf{U}}$ is the fluid velocity, $M$ is the mobility, and $\mu$ is the chemical potential derived from the free energy functional in Eq. (\ref{eq:free_energy}) by $\mu = \frac{\delta F}{\delta \psi} = A \psi + B \psi^3 - \kappa |\nabla^2 \psi|$. The fluid velocity is obtained by solving the Navier-Stokes equation. 

\begin{equation} \label{eq:NS-eqn}
\rho (\partial_{t} {\bf{U}} + {\bf{U}}\cdot \nabla {\bf{U}}) = - \nabla P + \eta \nabla^2 {\bf{U}}  - \psi \nabla \mu 
\end{equation}

where, the gradient of the hydrodynamic pressure $P$ is supplemented by the thermodynamic force $\psi\nabla\mu$. The simulations are carried out using a hybrid method, where $\psi$ is solved by finite difference and $\textbf{U}$ arises from lattice Boltzmann. For all the simulations reported in this work, the viscosities of both phases ($\psi = +1$ and $\psi = -1$) are maintained at a constant value of 0.625. The simulations are carried out under low Reynolds number with $\mathrm{Re} \sim \mathcal{O}(10^{-2})$. 

The capillary number $\mathrm{Ca} = \dfrac{\eta u_0}{\sigma_0}$ is calculated in two ways: first, by varying the interfacial tension $\sigma_0$ while keeping the self-propulsion velocity,  $u_0$ constant, and second, by varying $u_0$ while keeping $\sigma_0$ constant. In the first case, the parameters of the binary fluid interface, including the bulk free energy coefficients, are varied within the range $A = -B = -4.9 \times 10^{-3}$ to $ 0.019 $, and the interface penalty $\kappa$ is varied from $\mathrm{ 4.8 \times 10^{-3} }$ to $ 0.0189 $. This results, interfacial tension ($\sigma_0 = \sqrt{\frac{8\kappa |A|^3}{9B^2}}$), values in the range of $\sigma_0 \approx 4.6 \times 10^{-3} $ to $ 0.018$, while keeping the mobility $ M = 0.5 $ and velocity $u_0 = 8.9 \times 10^{-4}$ constant. In the second case, the interfacial tension is fixed at $ \sigma_0\approx 0.012 $, with $A = -B = 0.0133$ $\And$ $\kappa = 0.013$ while the propulsion velocity $u_0$ is incremented between $5.7 \times 10^{-4} $ and $0.0023$. The interfacial width $\left( \chi_{0} = \sqrt{\frac{2\kappa}{|A|}}\right)$ is fixed at a constant value of $1.4$ for all the reported results. The units of all the parameters used in this study are in lattice units (LU). We used open source lattice Boltzmann code Ludwig (version 0.21.0) to carry out simulations~\cite{kevinstratford_2024_12822477,ludwig_code}.

\section{Results and Discussion}

\begin{figure}
\centering
\includegraphics[width=1.0\textwidth]{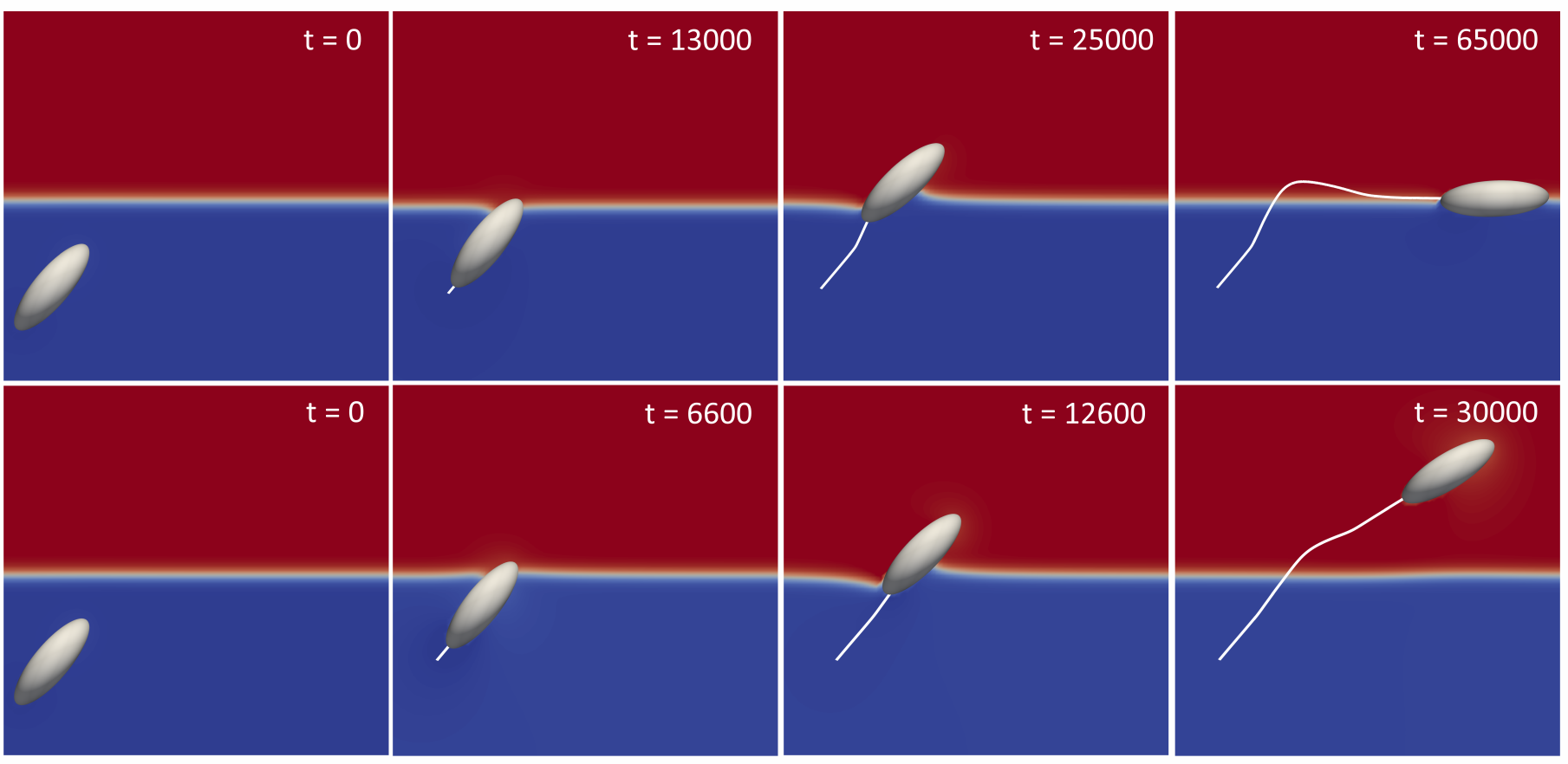}
  \caption{Snapshots showing a neutrally wetting microswimmer either being trapped or crossing a clean liquid-liquid interface, for $\mathrm{Ca}\approx 0.03$ (top row) and $\mathrm{Ca}\approx 0.12$ (bottom row) respectively. The initial projection angle is $\phi_0=50^\circ$ and interfacial tension $\sigma_0\approx 0.012$.}
\label{Snapshots}
\end{figure}

\begin{figure}
    \centering
    \includegraphics[width=1.0\textwidth]{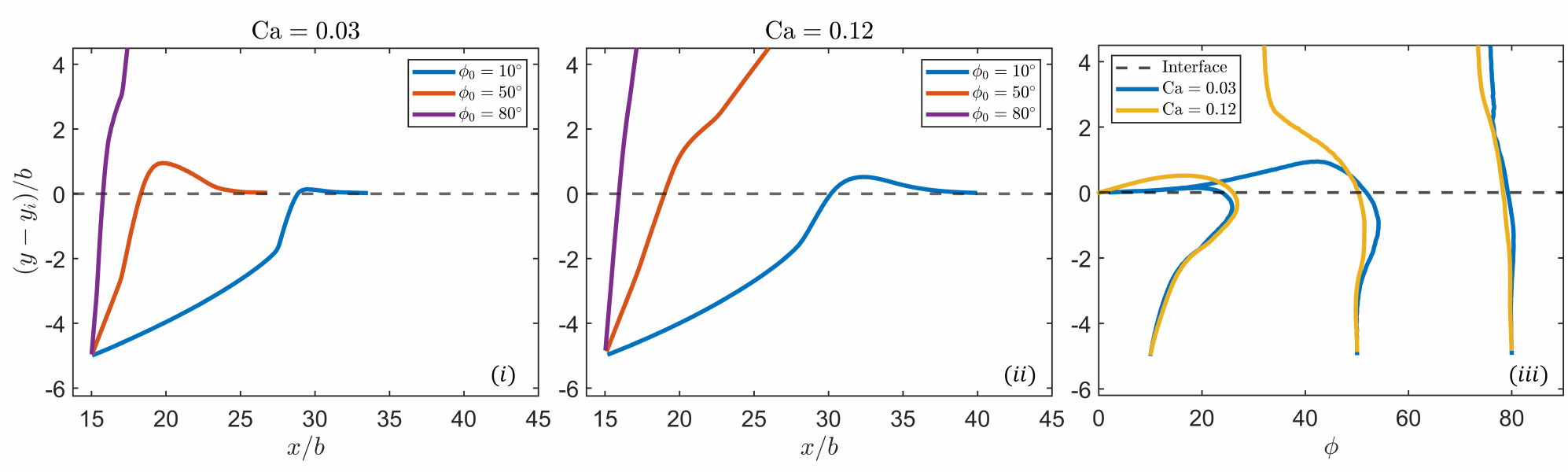}
    \caption{Trajectories in $x-y$ space  of a prolate swimmer for (i) $\mathrm{Ca} \approx 0.03$ and (ii) $\mathrm{Ca} \approx 0.12$ as well as (iii) in $\phi-y$ space, for initial alignment angles $\phi_0\approx  10^\circ$, $50^\circ$ and $80^\circ$, where $y_i$ is the interface position. The data corresponds to interfacial tension $\sigma_0 \approx 0.012$ and aspect ratio $AR=3$.}
    \label{traj_Constant_tension}
\end{figure}

{\color{black} Motivated by experimental observations~\cite{Cheon_SOFT_MATTTER_2024,Cheon2024_Arxiv_2024}, we study the dynamics of a rod-like prolate shaped swimming particle approaching a fluid-fluid interface at different $\mathrm{Ca}=\eta u_0/\sigma_0$ and initial projection angles $\phi_0$.} 
{\color{black} The particle was initialized at vertical distance $5b$ below the interface and its initial aligned angle $\phi_0$ was varied (Fig.~\ref{schematic}). 
The swimmers are neutrally wetted by both the fluids, corresponding to a contact angle $90^\circ$ at the interface, in agreement with the experiments~\cite{Cheon_SOFT_MATTTER_2024}. When a swimmer reaches an interface, its surface wets the fluids in continuous fashion (see intermediate times in (Fig.~\ref{Snapshots}). When the interfacial forces dominate, the particle is trapped, and eventually aligns parallel along the interface (see $\mathrm{Ca}\approx 0.03$ sample (top row) in (Fig.~\ref{Snapshots})). When the self-propulsion speed is increased, the swimmer can swim across the interface (see $\mathrm{Ca}\approx 0.12$ sample (bottom row) in Fig.~\ref{Snapshots}).

To study the dynamics, the particle trajectories were recorded across multiple capillary numbers and initial alignment angles $\phi_0$ (Fig.~\ref{traj_Constant_tension}).  When the swimmers are initially close to perpendicular to the interface, large $\mathrm{\phi_0}$, we observe a smooth crossing of the interface, while for small shallow angles the swimmers gets trapped at interface (Fig.~\ref{traj_Constant_tension}i). The critical angle $\phi_0$ when this happens, depends on the capillary number (Fig.~\ref{traj_Constant_tension}ii). When the $\mathrm{Ca}$ is reduced the, interfacial (Pickering) forces dominate over that active (swimming) forces and the swimmers are observed to be trapped at larger alignment angles (Fig.~\ref{traj_Constant_tension}i).} Additionally, reorientation of the swimmer occurs along its trajectory (Fig.~\ref{traj_Constant_tension}iii). Typically, the swimmer reorients towards the parallel of the interface. Interestingly, at shallow angles we observe initial reorientation towards the interface normal when the particle is well below the interface, followed by realignment parallel to the interface after the particle has been trapped (see {\it e.g.} $\phi_0 \approx 10^\circ$ curves in Fig.~\ref{traj_Constant_tension}iii). The initial reorientation likely arises from the far-field hydrodynamics. %Similar initial reorientation was also observed in experiments of Bacillus Subtilis~\cite{Cheon_SOFT_MATTTER_2024}.

{\color{black} To understand the trapping in more detail, we mapped a phase diagram as a function of the capillary number $\mathrm{Ca}$ and initial alignment angle $\phi_0$ (Fig. \ref{Prolate_phase}).} 
As it is mentioned in the previous section (Sec. \ref{Simulation details}) {\color{black} we considered two set of simulations: one where the capillary number $\mathrm{Ca}=u_0\eta/\sigma_0$ is calculated by varying the interfacial tension $\sigma_0$ for a constant velocity $u_0$ (blue symbols in Fig.~\ref{Prolate_phase}) and other by varying $u_0$ while keeping $\sigma_0$ constant (red symbols in Fig.~\ref{Prolate_phase}).
Our numerical phase diagram is compared against the experimentally observed swimming states~\cite{Cheon_SOFT_MATTTER_2024} (white symbols in Fig.~\ref{Prolate_phase}). The capillary number for the experiments, was calculated from the reported parameters, swimming speed (15--35~$\mu\mathrm{m\,s^{-1}}$), interfacial tension (10~$\mu\mathrm{N\,m^{-1}}$), and viscosity (30~$\mathrm{mPa\,s}$)~\cite{Cheon_SOFT_MATTTER_2024}, for Bacillus subtilis at the vicinity of isotropic-nematic interface.
 The simulation results match the experiments very well, and we observe very little difference between the two set of simulations (Fig.~\ref{Prolate_phase}).  

\begin{figure}
\centering
  \includegraphics[width=0.65\textwidth]{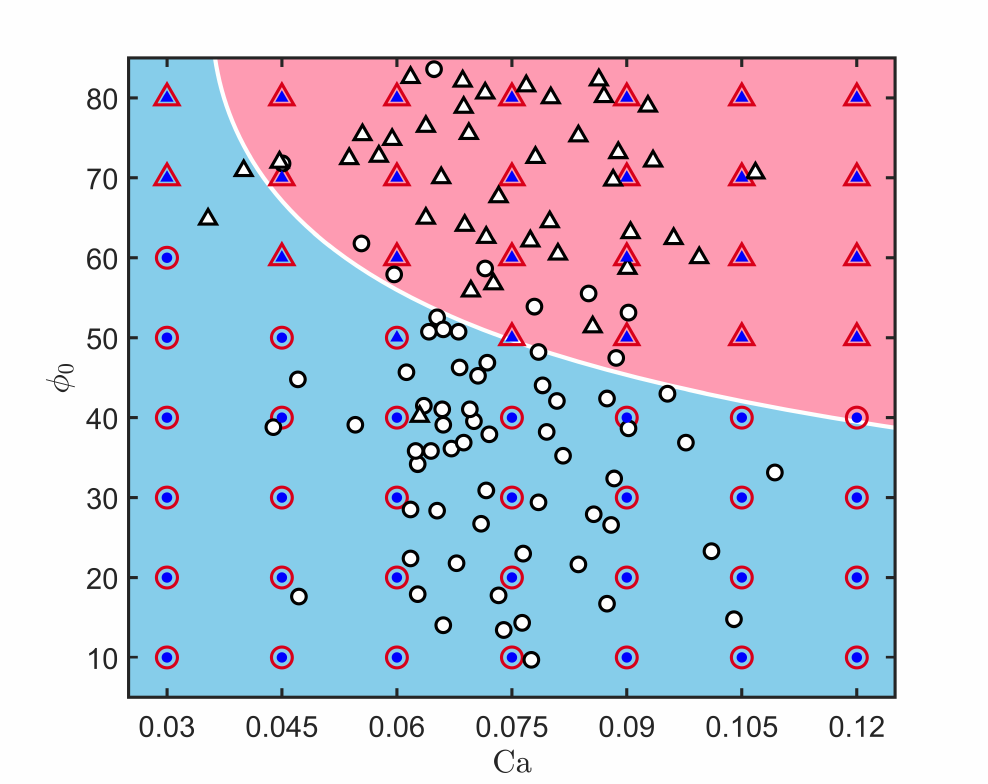}
  \caption{A state diagram of a prolate microswimmer showing crossing (triangles) and trapped (circles) states as a function of the capillary number $\mathrm{Ca}=u_0\eta/\sigma_0$ and initial alignment angle $\phi_0$.  The blue (red) symbols correspond to simulations where the swimming speed $u_0$ (interfacial tension $\sigma_0$) were varied. The open white symbols represent the experiments of Bacillus Subtilis crossing an isotropic-nematic interface~\cite{Cheon_SOFT_MATTTER_2024}. The solid white line represents a balance between active and interfacial energies (Eq.~\ref{Eq_Energy_balance}; see text for details).}
  \label{Prolate_phase}
\end{figure}

\begin{figure}[h!]
    \centering
    \includegraphics[width=1.0\textwidth]{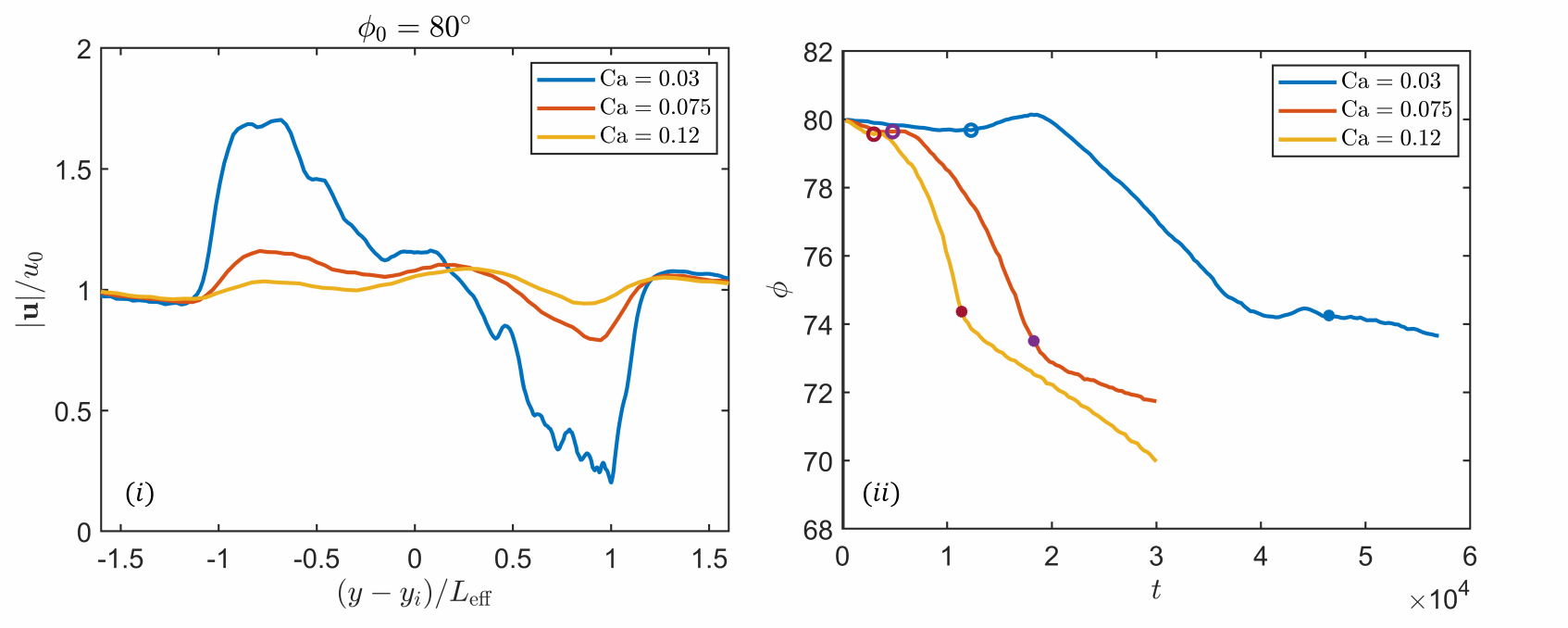}
    \caption{ (i) The normalized velocity shows the fluctuations when the prolate swimmer crosses the interface located at $y \approx y_i$. (ii) Temporal evolution of the orientation $\phi$ of the microswimmer during the interaction with the interface, where $\sigma_0 \approx 0.012$. The open and closed circles indicate the particle-interface interaction onset and offset points respectively. The onset and offset points are at $y_\approx y_i \mp L_\text{{eff}}$, below and above to the interface.}
\label{velocity_orientation}
\end{figure}

\subsection{Thermodynamic trapping} \label{Enegy_balance_model}

{\color{black} We hypothesize that the observed trapping/crossing dynamics arises from the balance between interfacial (trapping) and active forces. A spherical particle with a radius $R$ at height $h < R$ below a liquid-liquid interface eliminates a planar area $\pi\left (R^2 - h^2 \right )=\pi R^2\left(1-\cos\theta\right )^2$, where $\theta$ is related to the contact angle between the two fluids and the particle surface. Considering an interface with an interfacial tension $\sigma_0$, removing the particle into the lower fluid has an energetic cost $E=\pi R^2\sigma_0\left( 1 - \cos\theta\right )^2$~\cite{BINKS_LANGMUIR_2000}. For neutrally wetting particles, $\theta \approx 90^\circ$, as in the bacterial experiments~\cite{Cheon_SOFT_MATTTER_2024} and in our simulations, this gives a trapping energy $E=\pi R^2\sigma_0$.

We assume that this is the dominant trapping mechanism in our simulations. For a prolate swimmers, the area removed from the interface takes the shape of an ellipse and depends both the the aspect ration $a/b$ as well as the alignment angle, $\phi$~\cite{DAVIES_SOFTMATTER_2014}.
\begin{equation}
A_\text{rm}(\phi) = \frac{\pi a b^2}{\sqrt{b^2 \cos^2\phi + a^2 \sin^2\phi}}. 
\end{equation}
This gives an angle dependent maximum magnitude of trapping energy 
\begin{equation}
(\Delta E^p_{\text{th}})_\text{min} = A_\text{{rm}}(\phi)\sigma_0,
\end{equation}\label{eq:max}
where the thermodynamic trapping increases with decreasing $\phi$ and reaches the maximum when the rod-like particle is parallel to the interface $\phi\approx 0$. }

To predict the interface crossing behavior of the microswimmer, we equate two opposing energies (trapping and active). The trapping energy due to the interfacial thermodynamics is given as, $(\Delta E^p_{\text{th}})_\text{min} = -A_\text{{rm}}(\phi)\sigma_0$ which favors the trapping of the particle and active energy corresponding to the swimming. The latter can be defined as the product of translational drag coefficient ($\mathrm{\mu_{trans}}$), normal component of the swimming velocity ($u_0\sin\phi_0$) and the effective swimming length ($L_\text{{eff}}$), $\Delta E_\text{{active}} = \mu_\text{trans}u_0\sin\phi_0 L_\text{{eff}}$, where $\mathrm{\mu_\text{trans}= \frac{4\pi\eta a}{\ln\frac{2a}{b}-0.5}}$. The maximum trapping is observed when the center of the swimmer is at the interface. To escape the trapping potential posed by the interface, the swimmer needs to move an effective length perpendicular to the interface, $L_\text{eff}=\sqrt{a^2 \sin^2\phi_0 + b^2 \cos^2\phi_0}$. Now we balance both opposing energies as follows, 
\begin{equation}\label{eq_Enegy_model}
 \sigma_0 A_\text{rm} = \mu_\text{trans} L_\text{eff} u_0 \sin\phi_0.
\end{equation}
Reorganizing this, one can obtain an expression for the critical capillary number ($\mathrm{Ca^*}$).

\begin{equation}\label{Eq_Energy_balance}
    (4 \mathrm{Ca^*)}^{-1} = \frac{\sin\phi_0 (a^2 \sin^2\phi_0 + b^2 \\cos^2 \phi_0)}{b^2 (\ln(\frac{2a}{b})-0.5)}.
\end{equation}
The above expression gives the boundary between the crossing and trapped states (solid white line in Fig.~\ref{Prolate_phase}), and agrees very well both with the simulations and bacterial experiments~\cite{Cheon_SOFT_MATTTER_2024} (red/blue symbols and white symbols in Fig.~\ref{Prolate_phase}, respectively).}
% favoring the detachment of the particle from the interface  $\mathrm{E_\text{particle}}$.

\begin{figure}
\centering
\includegraphics[width=1.0\textwidth]{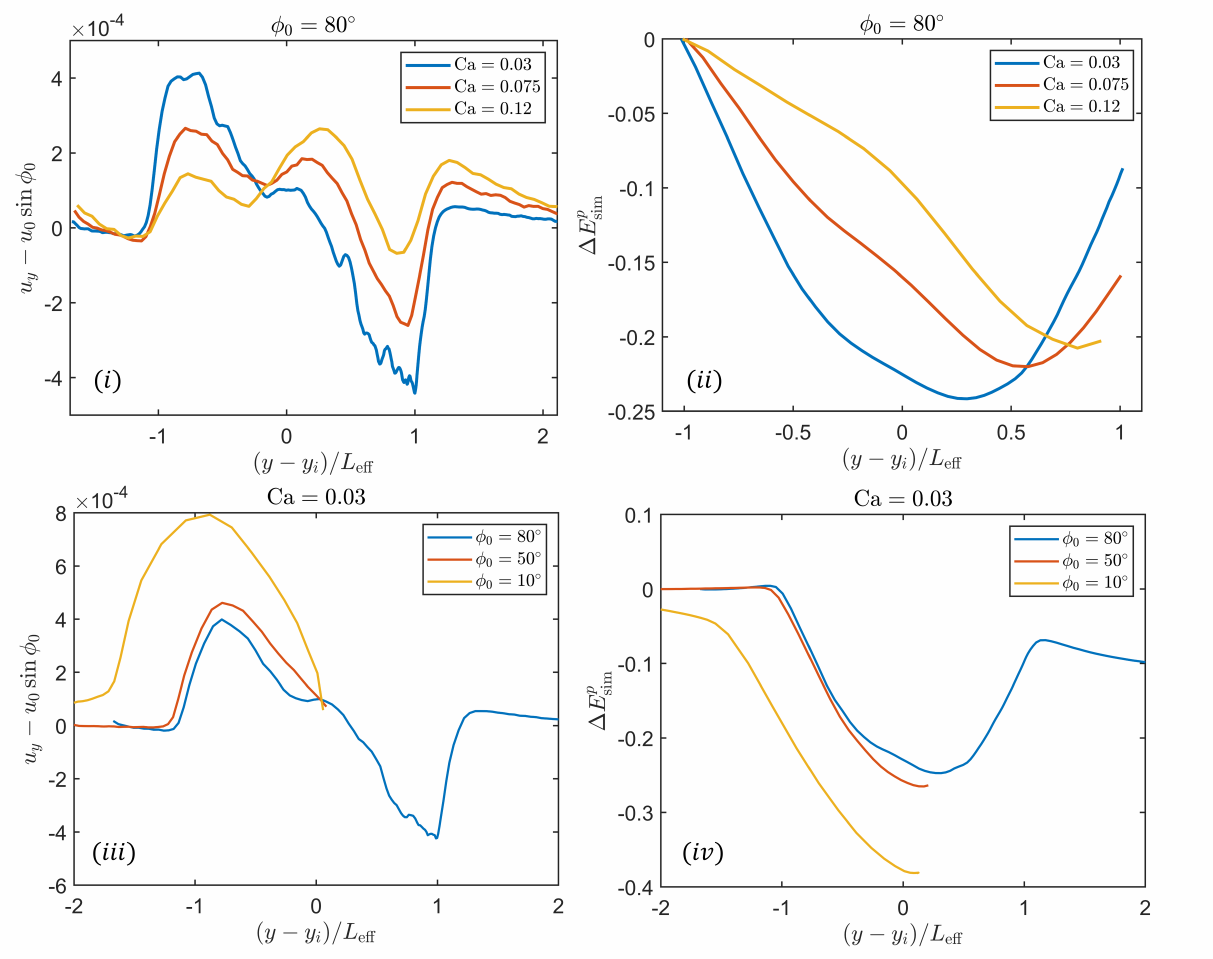}
  \caption{(i) The velocity component perpendicular to the interface $u_y - u_0 \sin\phi_0$ for $\mathrm{Ca}\approx 0.03,~0.075,~0.12$ at fixed initial alignment $\phi_0\approx 80^\circ$. (ii) The corresponding calculated spatial trapping potential $\Delta E^p_{\text{sim}}=-\mu_\text{trans}\int (u_y - u_0 \sin\phi_0)dy$. (iii) and (iv) $u_y - u_0 \sin\phi_0$ and $\Delta E^p_{\mathrm{sim}}$ for $\phi_0\approx 10^\circ, ~ 50^\circ, ~ 80^\circ$ at fixed $\mathrm{Ca}\approx 0.03$, respectively. The data corresponds to $\sigma_0 \approx 0.012.$} 
  \label{pinning_velocity_with_fit}
\end{figure}

\subsection{Particle dynamics}

The dynamics of swimmer is strongly influenced by the presence of liquid-liquid interface. During the early stages of the trajectory the swimmer speeds up at the onset of particle-interface interaction and then slows down until the complete separation from the interface (Fig \ref{velocity_orientation}i). This is consistent with a thermodynamic potential corresponding to a Pickering trapping of neutrally wetting particles at liquid-liquid interfaces~\cite{BINKS_LANGMUIR_2000}. When the Capillary number ($\mathrm{Ca}$) increases, the overall change in swimming velocity during interface crossing decreases. This behavior can be attributed to the increasing dominance of active force, which allow the particle to overcome the interfacial resistance smoothly. During this period, the translational motion of the microswimmer is accompanied by rotational motion (Fig \ref{velocity_orientation}ii). In a previous work of spherical squirmers~\cite{GIDITURI_PRF_2022}, it was observed that a pusher type  microswimmer reorients parallel to the interface solely based on the hydrodynamics. However, in the case of prolate shaped swimmer, rotational motion is resulted due to both hydrodynamics and thermodynamic interaction between the particle and the interface. It is interesting to note that the exit angle of the swimmer traversing the interface remains approximately same, whereas the rate of orientation increases with increasing capillary number $\mathrm{Ca}$ (Fig \ref{velocity_orientation}ii). This can be elucidated from the choice of $\beta = -1$ with increment in the $\mathrm{Ca}$, causing the swimmer to spend less time in the vicinity of the interface. The reorientation behavior is in agreement with the previous studies~\cite{GIDITURI_PRF_2022} where enhanced rate of orientation is due to an increase in the hydrodynamic for dipole $(d\phi/d t \sim B_2$), while the thermodynamic contribution to the reorientation remains constant for a constant interfacial tension $\sigma_0$.

{\color{black} In the absence of  perturbations, a single swimmer moves with a speed $u_0$ in a bulk fluid. To evaluate the effect of the presence of a liquid-liquid interface, we consider a velocity component $u_y(t) - u_0\sin\phi$ perpendicular to the interface (Fig.~\ref{pinning_velocity_with_fit}i). From the simulation data $u_y-u_0\sin\phi$ (Fig.~\ref{pinning_velocity_with_fit}i), a trapping energy can be estimated as, $\Delta E^p_{\text{sim}}=-\mu_\text{trans}\int (u_y - u_0 \sin\phi_0)dy$ (Fig.~\ref{pinning_velocity_with_fit}ii).  In an ideal situation, in the absence of hydrodynamic flows ($\mathrm{Ca} = 0$) one would expect a smooth parabolic potential with a minimum at the middle of the interface(i.e., $h=0$).

For reasonably small capillary numbers we observe an initial speed up of the vertical velocity, followed by smooth reduction of the speed (Fig.~\ref{pinning_velocity_with_fit}i). The integrated trapping potential is parabolic (see {\it e.g.} $\mathrm{Ca}\approx 0.03$ in Fig.~\ref{pinning_velocity_with_fit} ii). When the $\mathrm{Ca}$ is increased both the interface induced variations in the particle velocity become non-monotonic (see {\it e.g.} $\mathrm{Ca}\approx 0.075$ and $0.12$ in Fig.~\ref{pinning_velocity_with_fit}i). Also the position of the trapping potential minimum deviate (rightward shift) significantly from the ideal case (Fig. \ref{pinning_velocity_with_fit}ii). However, the maximum trapping energy, given by the potential minimum, remains more or less constant across the $\mathrm{Ca}$ considered, which implies a constant thermodynamic trapping due to a constant value of the interfacial tension and alignment angle $\phi$.

This observation is consistent with the theoretical thermodynamic trapping potential ($\Delta E_{\text{th}}^p$) of a prolate swimmer (Eq. ~\ref{Equation_dynamic_pinning_energy}), which can be defined in terms of the $A_\text{{rm}}$ as a function of the offset height $h$ and $\phi_0$
\begin{equation}\label{Equation_dynamic_pinning_energy}
    \Delta E_{\text{th}}^p = -\sigma_0 A_\text{rm}(h,\phi_0),
\end{equation}
where, 
\begin{equation}\label{Eq_Area_removed}
    A_\text{{rm}}(\phi_0,h) = \frac{\pi a b^2  [(a^2 sin^2\phi_0 + b^2 cos^2 \phi_0) -h^2]}{(a^2 sin^2\phi_0 + b^2 cos^2 \phi_0)^\frac{3}{2}}
\end{equation}
}

As the eq.~\ref{Equation_dynamic_pinning_energy} suggests, the potential energy depends on the alignment angle $\phi$, where the maximum trapping is predicted for swimmers parallel to the interface $\phi\approx 0$. Our simulations agree with this. For a constant value of capillary number, $\mathrm{Ca}\approx 0.03$, reduction in $\phi_0$ from angles close to interface normal ${\it{i.e.}}$ $90^\circ$ to a shallow angles $10^\circ$, a larger velocity component normal to the interface (Fig.~\ref{pinning_velocity_with_fit}iii) consequently a deeper potential is observed (Fig.~\ref{pinning_velocity_with_fit}iv).
\begin{figure}
    \centering
    \includegraphics[width=1.0\linewidth]{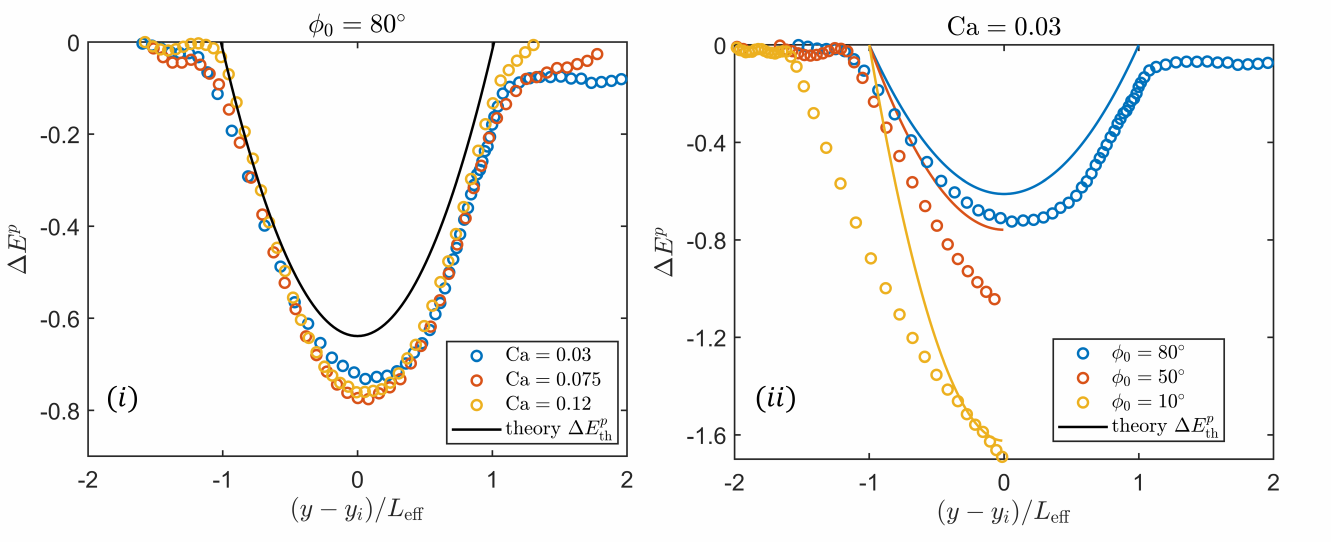}
    \caption{The trapping potential $\Delta E^p$ calculated from the binary fluid free energy (Eq.~\ref{eq:free_energy}; see text for details, where $\Delta E^p = F(\text{bulk})-F(\text{particle at interface})$). (i) $\Delta E^p$  as a function of the particle position $y$  for various capillary numbers $\mathrm{Ca}$ at fixed $\phi_0 = 80^\circ$. 
(ii) $\Delta E^p$ for  a range of $\phi_0$ at fixed $\mathrm{Ca} \approx 0.03$. The simulations were carried out with a constant interfacial tension $\sigma_0 \approx 0.012$.}  
    \label{pinning_potential_AR3}
\end{figure}

We can also calculate the evolution of the binary fluid free energy (Eq. \ref{eq:free_energy}) as a function of the particle’s distance, $y$, from the interface. Now the trapping potential $\Delta E^p$ can be extracted by calculating the difference in thermodynamic free energy when the swimmer is in the interface interaction region and the bulk region  (Fig.~\ref{pinning_potential_AR3}). Using this method, we obtained reasonable quantitative agreement between the theoretical  potential Eq.~\ref{Equation_dynamic_pinning_energy} (solid lines in Fig.~\ref{pinning_potential_AR3}). Comparing the simulations and theoretical results, it is observed that the effective range of swimmer–interface interaction is larger in simulations, especially for shallow angles, particularly at $\phi_0 = 10^\circ$. 
This discrepancy can be understood considering the phase trajectory ($y$ vs $\mathrm{\phi}$) of the swimmer (Fig.~\ref{traj_Constant_tension}iii). For the initial angle $\phi_0\approx 10$, it is observed  that below the interface swimmer is reorienting perpendicular to the interface reaching a maximum alignment angle $\phi\approx 30$. A similar reorienting behavior towards the interface normal is also reported in the bacterial experiments \cite{Cheon_SOFT_MATTTER_2024}.  This reorientation likely due to a hydrodynamic torque, $\tau_{\text{hydro}}$ arising from the coupling between the far-field flow-fields of the swimmer and the interface. The increased angle $\phi$ makes the apparent  $L_\text{{eff}}(t)>L_\text{{eff}}(\phi_0)$, thus the particle surface reaches the interface at distances $y>L_{\mathrm{eff}}(\phi_0)$. 

\begin{figure}
\centering
\includegraphics[width=1.0\textwidth]{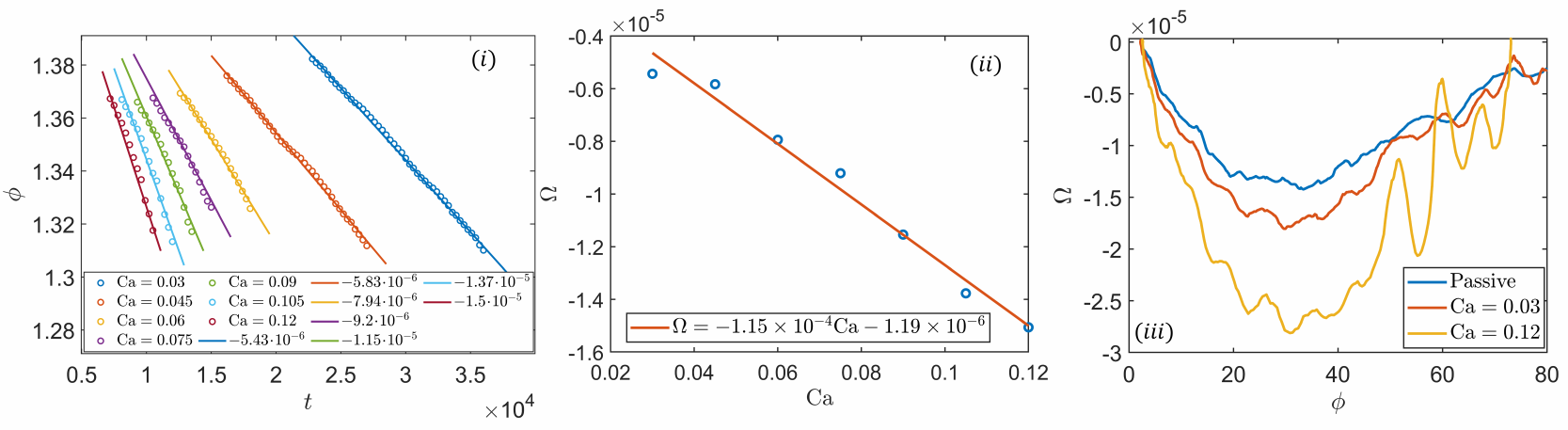}
  \caption{Rate of reorientation of a prolate swimmer interacting with a fluid–fluid interface with an initial alignment angle $\phi_0\approx 80^\circ$. (i) The temporal evolution of the orientation angle $\phi$  (in radians), the solid lines represent the linear fit, from which the slope ($\Omega$) has been obtained. (ii) The reorientation rate $\mathrm{\Omega} = \dfrac{d\phi}{dt}$ calculated from (i) as a function of the capillary number ($\mathrm{Ca}$).
 (iii) The reorientation rate $\Omega$ calculated for a passive ellipsoid (blue line) and two ellipsoidal shakers $B_1 =0, B_2\neq 0$ corresponding to $\mathrm{Ca}\approx 0.03$ and $0.12$ (see text for details). The simulations correspond to aspect ratio $AR=a/b = 3$ and interfacial tension $\sigma_0\approx 0.012$.} 
  \label{Reorientation_fig}
\end{figure}

The integrated potentials from the particle dynamics (Fig.~\ref{pinning_velocity_with_fit}) agree qualitatively with the potentials calculated from the free energy (Fig.~\ref{pinning_potential_AR3}). However, in the case of particle dynamics, the apparent maximum trapping energy is reduced. This is likely due to additional hydrodynamic interactions between the swimmer and the interface. It should also be noted, in the integration, a constant drag coefficient, $\mu_\text{trans}= \frac{4\pi\eta a}{\ln\frac{2a}{b}-0.5}$ was assumed. However, the drag coefficient itself likely depends on the distance and other details when the particle is near an interface~\cite{BONIELLO_PRE_2016,LOUDET_EPJE_2020,DANOV_CES_1995,DANOV_CES_1998}. 
Despite these discrepancies, all the data supports the idea of trapping the swimmers via Pickering effect.

Previous studies have proposed, that a spherical pusher-type microswimmer at an interface with no viscosity contrast tends to align parallel to the interface due to a hydrodynamic torque arising from the force dipole $B_2$ flows~\cite{GIDITURI_PRF_2022}.
We assume similar mechanism in our rod-like swimmers as well.  However, for a prolate-shaped particle a thermodynamic component of a torque, $\tau_{\text{thermo}}$ exists also, aligning the long axis of the ellipsoid along the interface. 
To evaluate the reorientation of the swimmers, we considered swimmers crossing the interface with an initial alignment angle $\phi_0\approx 80$ for various capillary numbers $\mathrm{Ca}\approx 0.03\ldots 0.12$ (Fig.~\ref{Reorientation_fig}), for a fixed interfacial tension $\sigma_0$. In all the case, the swimmers were observed to reorient towards the interface (Fig.~\ref{Reorientation_fig}i), with the reorientation rate increasing linearly the with capillary number (Fig.~\ref{Reorientation_fig}ii). To study the reorientation in more detail,  passive particle and two shakers ($B_1=0$; $B_2\neq 0$) corresponding to $\mathrm{Ca}=\eta B_2/\sigma_0\approx 0.03$ and $0.12$ were considered (Fig.~\ref{Reorientation_fig}iii). In all the cases, the particles were initialised at the middle of the interface with an initial alignment $\phi_0\approx 80$. The results agree qualitatively what is expected. The highest $B_2$ corresponding to the largest $\mathrm{Ca}$ is observed to rotate fastest, while the passive ellipsoid is the slowest. {\color{black} This suggests a reorientation arising from a torque $\tau$ with both thermodynamic and hydrodynamic components $\tau = \tau_{\mathrm{thermo}} + \tau_{\mathrm{hydro}}$, where $\tau_{\mathrm{hydro}}\sim B_2$ and $\tau_{\mathrm{thermo}}\sim  -\tfrac{\partial \Delta E^p}{\partial \phi.}$.}

%\subsection{Details concerning the comparison with the bacterial experiments}

\subsection{Conclusions}
We have investigated the dynamics of a prolate shaped microswimmer approaching a liquid-liquid interface for a range of capillary numbers, $\mathrm{Ca}$ and initial orientation angles, $\phi_0$. We found two swimming states (trapped and crossing) in the $\mathrm{Ca}$ and $\phi_0$ space. The trapping dynamics of the swimmer can be understood by balancing active and interfacial forces. 

Our simulation results agree well with the experimental findings of Cheon \textit{et al.}\cite{Cheon_SOFT_MATTTER_2024} for rod-like bacteria crossing an interface of isotropic-nematic coexistence.  To compare to the bacterial experiments, we converted the data given in Fig.~2 in~\cite{Cheon_SOFT_MATTTER_2024} using the supplementary figure S8 as well as reported values for viscosity $\eta \approx 30\mathrm{mPas}$ and interfacial tension $\sigma_0\approx 10\mu\mathrm{Nm}^{-1}$~\cite{Cheon_SOFT_MATTTER_2024}. 
In the experiments~\cite{Cheon_SOFT_MATTTER_2024}, the trapping dynamics was explained by a spring like force arising from the interfacial deformations. This is different to our model. We do not observe such a deformation, but the swimmer surface is wetted by the two fluids (see {\it e.g.} Fig.~\ref{Snapshots}).  In our study, we propose a parabolic trapping potential arising from the Pickering effect of neutrally wetting particles, {\color{black} which qualitatively agree with a trapping force measured in experiments of DEX-PEG interfaces~\cite{Cheon2024_Arxiv_2024}.}

 Our result also shows that the translational motion of the swimmer is accompanied by a rotational motion. The degree of reorientation depends on the hydrodynamic and thermodynamic torques, which both rotate the swimmer parallel to the interface. The hydrodynamic torque that arises from the force dipole $B_2$ term increases linearly with $\mathrm{Ca}$ for $\beta\approx -1$ swimmers, whereas the thermodynamic torques remain constant for a fixed aspect ratio $AR$.

\begin{acknowledgments}
\noindent The authors gratefully acknowledge the High Performance Computing (HPC) support provided by AGASTYA at IIT Jammu, India, and CURTA at MCIA, France, for facilitating the computational resources used in this work.
\end{acknowledgments}
\nocite{*}

\bibliography{apssamp}% Produces the bibliography via BibTeX.

\end{document}